\documentclass{PoS}

\title{Modification of Hadron Structure and Properties in Medium}

\ShortTitle{Hadron Structure}

\author{\speaker{A.~W.~Thomas} \\ 
        CoEPP and CSSM, School of Chemistry and Physics, 
        University of Adelaide, SA 5005 Australia}

\abstract{
In the quest to understand QCD there are a number of outstanding challenges.
Here we focus on one of these, namely what one expects to happen to the 
structure of a hadron when it is immersed in a nuclear medium. We argue 
that the necessary changes in the quark structure are intimately related 
to nuclear binding and saturation. Some of the potential experimental 
implications of these ideas are discussed. 
          }

\FullConference{XV International Conference on Hadron Spectroscopy-Hadron 2013\\
		4-8 November 2013\\
		Nara, Japan }

\begin{document}

\section{Introduction}
Over the past decade our capacity to calculate hadron properties from 
lattice QCD has improved dramatically. Ground state masses are well 
under control~\cite{Durr:2008zz} 
and there has been substantial progress with respect to 
electromagnetic form factors~\cite{Collins:2011mk} and 
even GPDs~\cite{Alexandrou:2011nr}. The especially challenging 
problem of disconnected diagrams, which for example are entirely responsible 
for the strange quark contributions 
to nucleon properties~\cite{Leinweber:1999nf}, have also seen some 
remarkable success~\cite{Leinweber:2004tc,Leinweber:2006ug,Young:2006jc}. 
The relatively accurate determination 
of the strange sigma commutator~\cite{Young:2009zb,Shanahan:2012wh} 
has had unexpectedly dramatic 
consequences for the accuracy with which one can calculate the cross 
sections for some particularly promising dark 
matter candidates~\cite{Giedt:2009mr}.

The study of hadron properties along the quark mass 
axis~\cite{Detmold:2001hq}, which of course is 
not available experimentally, has given us important new insights into how
hadron structure works. For example, we have discovered that meson cloud 
effects die fairly rapidly once the current quark masses rise above about 
50 MeV (or equivalently the pion mass 
above 0.4 GeV)~\cite{Young:2002ib}. 
We eagerly await the exploitation of this through the development 
of constituent quark models in this mass region, where they can be 
expected to work well; with the connection to data made after chiral 
extrapolation~\cite{Cloet:2002eg}.

Tackling resonance properties is a much trickier proposition, with the 
work of Luescher~\cite{Luscher:1990ux} providing a 
critical framework within which one 
can calculate two-body phase shifts. We are particularly excited by 
a new approach in which an appropriate Hamiltonian model is tuned to 
lattice data on a finite volume and then the physical resonance 
properties deduced from an infinite volume calculation. For an 
introduction to the method in the simple case of the $\Delta$ baryon 
we refer to Ref.~\cite{Hall:2013qba}.

Many baryon systems present far greater challenges. There have been 
some surprises in the case of di-baryons, with the infamous $H$ seeming 
to be still just on the edge of 
discovery~\cite{Inoue:2010es,Beane:2011iw,Shanahan:2011su} 
as the lower limit on its 
mass continues to rise~\cite{Carames:2013hla}. 
In retrospect, the insights into the role 
of the meson cloud that have come from studying baryon properties  
versus light quark mass provide even stronger support for the physics 
included in the original study of 
Mulders and Thomas~\cite{Mulders:1982da}. There, in contrast 
with the work of Jaffe~\cite{Jaffe:1976yi}, 
meson cloud corrections were included {\em but} 
in a way that recognised they would be suppressed strongly as the 
size of the hadron increased (roughly as one over the cube of the 
hadron size). This was why Mulders and Thomas found the $H$ to be around 
the $\Lambda$-$\Lambda$ threshold, rather than deeply bound -- the 
attractive pionic self-energy terms for the $H$ were suppressed 
with respect to the two $\Lambda$'s because of the former's significantly 
larger radius. This alone may well explain the otherwise surprising 
absence of most low mass exotic objects. Although we do note that 
the possible strangeness minus one 
state reported by the FINUDA Collaboration~\cite{Agnello:2005qj} 
at Frascati is remarkably close to 
the corresponding mass calculated in Ref.~\cite{Mulders:1982da}. 	 

When one thinks of atomic nuclei within the framework of QCD it is 
difficult to imagine how the complex many-body structure of a free 
nucleon cannot be modified by the presence of other hadrons separated 
by distances similar to the size of the hadrons themselves. For example, 
this may lead to the exchange of quarks between neighbouring colorless 
clusters or even the formation of hidden color configurations. The 
reliable calculation of such effects presents a formidable theoretical 
challenge.

A related effect, which is rather more amenable to modelling, is the effect 
of the strong scalar mean fields that we have known for more than 40 years 
play a major role in nuclear structure. In particular, since the development 
of the Paris potential~\cite{Lacombe:1980dr} 
on the basis of dispersion theory one has known that 
the attractive intermediate range NN force is an isoscalar, Lorentz scalar 
force associated with two-pion exchange, often represented through the 
exchange of a $\sigma$ meson. In a relativistic mean-field model such 
as Quantum Hadro-dynamics (QHD)~\cite{Serot:1984ey}
this led to a mean scalar field in a heavy 
nucleus of order 0.5 GeV -- more than half the mass of the nucleon itself. 
it is inconceivable that the application of such a field would not lead to 
significant modification of the properties of a bound nucleon.

In Sect. 2 we briefly review the quark-meson coupling (QMC) model which was 
developed by Guichon and 
colleagues~\cite{Guichon:1987jp,Guichon:1995ue,Saito:2005rv} 
to tackle this issue. We shall 
see that this model very naturally explains the saturation of nuclear binding 
and provides very helpful guidance as to the possible experimental 
consequences. In Sect. 3 we recall the extension of the QMC model using 
NJL, rather than the MIT bag model. Because the model is covariant it 
provides a more reliable framework for discussing changes in the
structure functions of bound nucleons. The final section contains some 
concluding remarks.

\section{QMC}
Within this model the valence quarks in the bound nucleon  
interact directly with the scalar and vector 
mesons that give rise to the mean fields in a nuclear medium. In its 
most natural form the isoscalar vector meson, the $\omega$, simply 
redefines energy levels and in uniform matter has no dynamical effect. 
On the other hand, the scalar field modifies the Dirac equation for the 
confined quark, enhancing the lower Dirac component of the 1s wave 
function as the density rises. Since the scalar coupling to the nucleon as 
a whole depends on the integral over the upper component squared minus the 
lower component squared, the effective $\sigma$-nucleon coupling decreases with 
increasing density. This in turn constitutes a new saturation mechanism 
for nuclear matter and leads to considerably lower scalar mean-fields 
than those appearing in QHD. 

Physically we may interpret this tendency for the scalar coupling to 
decrease as a response of the internal structure of the nucleon to 
the applied mean scalar field. By analogy with the more familiar 
electric and magnetic polarizabilities, this is known as the 
scalar polarizability of the nucleon~\cite{Guichon:1987jp,Chanfray:2003rs}. 
An initial attempt to derive this term 
from lattice QCD proved quite promising~\cite{Thomas:2004iw} 
but this needs further investigation. 

Starting with QMC one can derive an equivalent energy functional and hence 
extract Skyrme forces equivalent in this sense to the original model. 
The first investigation of this kind revealed many-body forces that 
were a direct consequence of the 
scalar polarizability~\cite{Guichon:2004xg}, while a later 
study led to density dependent Skyrme forces that produced rather 
good results for closed-shell, finite 
nuclei~\cite{Guichon:2006er}. Once again the origin
of the density dependence was the scalar polarizability, which modifies 
the force between two nucleons because of the presence of others. It 
will be very important to see further applications of these derived 
forces. It is particularly interesting that in a study of more than 
200 phenomenological Skyrme forces, many with 10-20 parameters fitted 
to various nuclear data, the forces derived from the QMC model with just 
3 parameters performed extremely well~\cite{Dutra:2012mb}.

Of course, if we were just concerned with nuclear binding energies 
the model would be of limited interest. However, because of the change 
of the valence quark structure, we expect that every property of the 
nucleon must change with density. Particular attention has been paid 
to the modification of the elastic form factors of the nucleon for 
which there has been an ambitious program at 
JLab~\cite{Strauch:2012nra,Malace:2010ft,Paolone:2010qc,Strauch:2002wu}. The 
QMC predictions~\cite{Lu:1997mu,Lu:1998tn}, 
made more than a decade before the actual measurements,   
anticipated key features of the data. Unfortunately, a multitude of 
unknown parameters associated with the more conventional analysis of the 
data make it difficult to draw a firm conclusion at present. 

A modification of the proton electric form factor has long been recognised 
as a potential correction to the longitudinal response of a nucleus in 
quasi-elastic electron scattering. There we eagerly await results from the 
extensive study conducted a few years ago at Jefferson Lab. A complete 
modern calculation based on a model such as QMC is urgently needed to 
complement that measurement. 

Deep inelastic nuclear scattering from nuclei has been of enormous interest 
since the discovery of the EMC effect in the early 80's at CERN. Early 
calculations within QMC showed that, indeed, the modifications of the 
structure of the bound nucleon predicted within that model were capable 
of producing just such an effect~\cite{Saito:1992rm}. 
However, the static bag model, 
upon which QMC is based, is not ideal for such calculations and more 
recent work has focussed on a similar approach within the NJL model.

\section{NJL Model for Nuclear Matter}
Starting with the NJL model one can construct nucleons and the mesons 
that bind them, creating a covariant, quark-level description of 
atomic nuclei where the Faddeev wave function of the bound 
nucleon is self-consistently modified in-medium~\cite{Bentz:2001vc}. 
This model not 
only generates remarkably good structure functions for the free 
nucleon~\cite{Cloet:2007em,Cloet:2005pp}
but also reproduces the EMC data for nuclei across 
the periodic table~\cite{Cloet:2006bq}. 
The challenge is how to test the model in ways that have not yet 
been measured. 

In this regard there have been two very important suggestions. The first 
relates to the importance of the lower Dirac component of the valence 
quark wave function in spin structure functions. Calculations of the 
EMC effect on the structure function of a bound nucleon 
suggest that it should be expected to be as much as 
twice as large as the unpolarised EMC effect~\cite{Cloet:2006bq}. 
Such experiments must be 
pursued at JLab after the 12 GeV upgrade is complete. 

The second test of this new paradigm for nuclear structure involves 
the so-called iso-vector EMC effect~\cite{Cloet:2009qs}. 
This was discovered in connection with 
the NuTeV anomaly, where it accounts for more than 1$\sigma$ of the nominal 
deviation from the Standard Model. The idea is very simple, namely 
in a nucleus with more neutrons than protons there is an iso-vector mean-field 
which modifies the structure function of {\em every nucleon}. Thus simply 
subtracting the free structure functions of the extra neutrons does not 
eliminate their effect. A number of experiments have been proposed 
which should reveal this effect. 
These include parity violating DIS~\cite{Cloet:2012td}, 
again ideally suited to JLab at 12 GeV, and even better charged current 
DIS on nuclei at a future electron-ion collider~\cite{Thomas:2009ei}.

\section{Conclusion}
Taking seriously QCD as the fundamental theory of the strong interaction, 
with quarks and gluons as the fundamental degrees of freedom, leads to 
a new paradigm for nuclear structure. It implies a new mechanism for 
nuclear saturation and predicts significant changes in the structure 
of the clusters of quarks with nucleon quantum numbers that occupy shell 
model orbits. It is imperative that the predictions of the observable 
consequences of these ideas be pursued viorously in the next few 
years.

\end{document}